\documentclass[12pt]{article}

\usepackage{amssymb}
\usepackage[dvips]{graphicx}
\usepackage{amscd}

\def\dj{\hbox{d\kern-0.347em \vrule width 0.3em height 1.252ex depth
-1.21ex \kern 0.051em}}

\begin{document}

\setlength{\oddsidemargin}{0cm}
\setlength{\baselineskip}{7mm}


\thispagestyle{empty}
\setcounter{page}{0}

\begin{flushright}
CERN-PH-TH/2006-243  \\
IFT-UAM/CSIC-06-60    \\
{\tt hep-th/0611312}
\end{flushright}

\vspace*{1cm}

\begin{center}
{\bf \Large Scaling Phenomena in Gravity from QCD}

\vspace*{1cm}

Luis \'Alvarez-Gaum\'e,$^{\,\rm a,}$\footnote{E-mail: 
{\tt Luis.Alvarez-Gaume@cern.ch}}
C\'esar G\'omez$^{\rm\, a,b,}$\footnote{E-mail: 
{\tt Cesar.Gomez@uam.es}} and 
Miguel A. V\'azquez-Mozo$^{\rm\, a,c,}$\footnote{E-mail: 
{\tt Miguel.Vazquez-Mozo@cern.ch}}

\vspace*{0.25cm}

\begin{quote}

$^{\rm a}$\,\,{\sl Theory Group, Physics Department, CERN, CH-1211
Geneva 23, Switzerland}

$^{\rm b}$\,\,{\sl Instituto de F\'{\i}sica Te\'orica UAM/CSIC, 
Universidad Aut\'onoma de Madrid, E-28049 Madrid, Spain}

$^{\rm c}$\,\, {\sl Departamento de F\'{\i}sica Fundamental, Universidad de 
Salamanca, Plaza de la Merced s/n, E-37008 Salamanca, Spain} and {\sl
Instituto Universitario de F\'{\i}sica Fundamental y Matem\'aticas (IUFFyM),
Universidad de Salamanca, Spain}

\end{quote}

\vspace*{1cm}
{\bf \large Abstract}
\end{center}

\noindent
We present holographic arguments to predict properties of strongly
coupled gravitational systems in terms of weakly coupled gauge
theories.  In particular we relate the latest computed value for the
Choptuik critical exponent in black hole formation in five dimensions,
$\gamma_{\rm 5D}=0.412 \,\pm 1\%$, to the saturation exponent of
four-dimensional Yang-Mills theory in the Regge limit, $\gamma_{\rm
BFKL} \simeq 0.410$.

\vspace*{2cm}

\centerline{\sl To Pere Pascual, in memoriam}

\newpage

\section{Introduction}

The realization of the holographic principle \cite{holography} in
terms of a duality between gravity and a conformal field theory
\cite{ads/cft} allows, in principle, the study of gauge and gravitational
dynamics beyond their perturbative regimes. However, most of the
applications of the gauge/gravity correspondence so far have dealt
with extracting gauge dynamics at large 't Hooft coupling from weak
gravity. This is because the study of strong gravity effects involves
backgrounds where the curvature becomes of the order of the string
scale at some point. Thus, the worldsheet dynamics of the
corresponding string theory becomes strongly coupled. This makes it
difficult to obtain reliable results to be compared with the gauge
theory calculations.

Because of this, it is of interest to find a window of gravitational
phenomena that, using holography, would allow a description in terms
of weakly coupled gauge dynamics. Generically, critical behavior is a
very robust physical phenomenon and independent of many details of the
dynamics involved.

Some years ago \cite{choptuikPRL} the existence of critical behavior in
black hole formation was discovered in numerical simulations of
gravitational collapse of a free massless scalar field (see
\cite{reviews_choptuik} for a review). Consider a family 
of scalar field initial conditions labeled by a real number $p$ such
that for small values of this parameter the collapse of this initial
configuration leaves behind flat Minkowski space-time, while for large
$p$ a black hole is formed. Choptuik \cite{choptuikPRL} found that
there is a critical value $p^{*}$ such that a black hole is formed
when $p>p^{*}$ while the field disperses if $p<p^{*}$. Moreover, for
supercritical initial conditions close to $p^{*}$ the radius of the
black hole horizon $r_{0}$ scales as
\begin{eqnarray}
r_{0} \sim (p-p^{*})^{\gamma},
\label{scaling}
\end{eqnarray}
where the critical exponent $\gamma$ is independent of the particular
family of initial conditions chosen. Numerical simulations showed that
in four dimensions $\gamma\approx 0.372$. This scaling behavior is
associated with departures from linear evolution when considering
initial data close to the critical solution. In $D$ space-time
dimensions the black hole mass is related to the radius of the horizon
by $M_{\rm BH}\sim r_{0}^{D-3}$. Thus, the critical scaling of this
quantity is given by $M_{\rm BH}\sim (p-p^{*})^{(D-3)\gamma}$.

We believe this is precisely the type of strong gravitational
phenomenon accessible, in principle, to a gauge theory calculation at
weak coupling. Next we have to identify the right gauge theory
framework.  For this we take into account that the production of black
holes at threshold involves strong gravitational fields as well as
large velocities.  This suggest the Regge domain of large $s$ but weak
coupling in the gauge dual. This is dominated by the exchange of a
BFKL pomeron \cite{pomeron} (see \cite{books} for review).  

It is known that in the one-pomeron approximation the total cross
section grows with the energy as $\sigma_{\rm tot}(s) \sim
s^{\alpha(0)}$, where $\alpha(0)>1$ is the intercept of the Regge
trajectory. Due to the power-like growth with $s$, this amplitude
eventually violates the Froissart-Martin unitarity bound, $\sigma_{\rm
tot}(s)\leq {\pi\over m_{\pi}^{2}} (\log{s})^{2}$. It is expected that
when the energy increases, saturation effects inside the hadron lead
to a departure from the linear evolution for the BFKL kernel.  This
leads to a decrease of the total cross section making it compatible
with unitarity requirements.

We conjecture that these two phenomena, the onset of nonlinear effects
in critical gravitational collapse and in the evolution of the BFKL
kernel, are dual and lead to the same critical exponents. Namely, we
propose that the saturation threshold in pomeron physics is a
holographic description of (``zero mass'') black hole formation in
gravitational collapse in five dimensions. The string regime we
describe corresponds to curvatures of the order of the string scale
but with small string coupling constant. In this sense the BFKL
pomeron at weak coupling provides the dual description of the strongly
coupled worldsheet string dynamics.  A BFKL calculation of the
critical exponent gives a result in surprisingly good agreement with
numerical simulations of five-dimensional gravity 
\begin{eqnarray}
\left.
\begin{array}{cc}
\gamma_{\rm 5D}=0.408\,\pm 2\% & \cite{sorkin_oren} \\
\gamma_{\rm 5D}=0.412\,\pm 1\% & \cite{bland}
\end{array}
\right\},
\hspace*{1cm} 
\gamma_{\rm BFKL}=0.409552.
\label{result}
\end{eqnarray}
In the last section we present some heuristic arguments supporting this
result.

Recently, a holographic interpretation of the pomeron was proposed in
\cite{BPST} according to which the BFKL Hamiltonian is identified with
the Laplacian for a type of metric perturbation in the
five-dimensional gravitational dual (see also \cite{others} for other
studies of the QCD pomeron in the context of the AdS/CFT
correspondence). Using this duality they obtained an expression for
the BFKL eigenvalue $\chi_{0}(\nu)$ in the strong 't Hooft coupling
regime, $g^{2}N\gg 1$. Notice, however, that our aim here is not to
use gravity to obtain results for QCD at strong coupling. Rather, we
identify phenomena in perturbative QCD providing reliable information
on strong gravitational effects in five dimensions. Although one cannot
exclude the possibility that we are dealing with a numerical coincidence, 
we are tempted to believe that this is unlikely,
given the wildly different mathematical and physical tools leading to the remarkable
agreement shown in Eq. (\ref{result}).  On the contrary, we think then that our 
arguments might provide an evidence for the validity of the holographic hypothesis in a region
hitherto unexplored.

\section{Criticality in black hole formation}

Numerical simulations of the gravitational collapse of a family of
scalar field configurations show the appearence of a scaling law for
the radius of the black hole formed. This implies the existence, in
infinite dimensional phase space, of a codimension one critical
hypersurface separating initial data leading to black hole formation
from those where the scalar field disperses leaving behind flat space
at late times.

Interestingly, the critical solution acting as an attractor for all
initial data at the critical surface has interesting properties
\cite{choptuikPRL}. Physically it can be pictured as describing the
formation of a black hole at ``zero mass'' and therefore presents a
naked singularity at the origin. Moreover, this critical solution has
the property of discrete self-similarity (DSS). This means that if we
denote by $Z_{*}(t,r)$ any of the components of the metric or the
scalar field, one finds the following symmetry
\begin{eqnarray}
Z_{*}(t,r)=Z_{*}(e^{\Delta}t,e^{\Delta}r).
\end{eqnarray}
From the numerical analysis the ``echoing'' period
$\Delta$ is extracted to be $\Delta\approx 3.44$. Here we take
the convention that the black hole forms as $t\rightarrow 0^{-}$. 

For initial conditions slightly away from the critical surface the
critical solution $Z_{*}(t,r)$ acts as a transient attractor. The
dynamical evolution drives the system to an ``echoing regime'' where
the metric is close to the critical one and approximately self-similar
near the origin. Eventually, the solution is repelled from the
critical surface to evolve either to Minkowski space-time or to form a
black hole.

The scaling in the black hole horizon can be understood as a
consequence of the dynamical instability of the critical solution,
i.e. the existence of a repulsive direction in phase space that takes
the metric away from the critical one \cite{gundlach_und}. Indeed,
introducing a fiducial length scale $\ell_{0}$ it is possible to
define coordinates $(\tau,\zeta)$ by
\begin{eqnarray}
\tau\equiv \log\left(-{t\over \ell_{0}}\right), \hspace*{1cm}
\zeta\equiv \log\left(-{r\over t}\right)-\xi_{0}(\tau),
\end{eqnarray}
where $\xi_{0}(\tau)$ is periodic with period $\Delta$. DSS acts now
by discrete translations in the $\tau$ coordinate,
$(\tau,\zeta)\rightarrow (\tau+\Delta,\zeta)$. Notice that in
this coordinates the black hole forms at $\tau\rightarrow -\infty$.

For slightly supercritical solutions with $p\gtrsim p^{*}$ we can
write $Z_{p}(\tau,\zeta)$ as a perturbation of the
critical solution
\begin{eqnarray}
Z_{p}(\tau,\zeta)\sim
Z_{*}(\tau,\zeta)+\alpha_{1}(\tau,\zeta)(p-p^{*})e^{\lambda_{1}\tau}
+\ldots
\end{eqnarray}
$\lambda_{1}<0$ is the eigenvalue associated with the repulsive
direction and the dots stand for the terms associated with the other
(positive) eigenvalues whose contribution is exponentially suppressed
for large negative $\tau$. The existence of a growing mode leads to a
departure from the linear analysis. This happens for a value
$\tau_{*}(p)$ of the (dimensionless) $\tau$ coordinate which in turn
defines the scale
\begin{eqnarray}
t_{*}(p)=\ell_{0}e^{\tau_{*}(p)}\sim \ell_{0}(p-p^{*})^{-{1\over
\lambda_{1}}}.
\end{eqnarray}
As explained in \cite{gundlach_und} $t_{*}(p)$ is the only
dimensionful quantity in the problem. It sets the scale of the black
hole apparent horizon and as a consequence the scaling (\ref{scaling})
is obtained.  The critical exponent is determined by the negative
eigenvalue $\lambda_{1}$ by 
\begin{eqnarray}
\gamma=-{1\over \lambda_{1}}.
\end{eqnarray}

A more detailed analysis \cite{gundlach_und,hod_piran} shows that in the case
of critical solutions with DSS there are periodic wiggles superimposed to the
scaling law (\ref{scaling})
\begin{eqnarray}
r_{0} \sim (p-p_{*})^{\gamma} e^{f[\log{(p-p^{*})}]},
\end{eqnarray}
where $f(x)$ is a periodic function. On general grounds it can be
proved that the period of this function is
${\Delta/\gamma}$. Depending, however, on the particular model under
study shorter periodicities compatible with this one are possible. For
example, in the critical collapse of a massless scalar field the
period of the wiggles is halved to ${\Delta/(2\gamma)}$
\cite{gundlach_und,hod_piran}.

The analysis presented here is completely general. In
the collapse of a massless scalar field both the critical exponent for
the black hole size $\gamma$ and the period of the echo
$\Delta$ have been computed in several dimensions
\cite{sorkin_oren,bland}. Specially interesting for our later discussion are
the results for five-dimensional gravity which, as mentioned above,
give a value $\gamma_{\rm 5D}= 0.408\,\pm 2\%$ for the critical
exponent with an echoing period $\Delta_{\rm 5D}= 3.19
\pm 2\%$ \cite{sorkin_oren} (the values found in Ref. \cite{bland} are
$\gamma_{\rm 5D}=0.412\,\pm 1\%$ and $\Delta_{\rm 5D}=3.10\pm 0.1$).
In the following section we show how a QCD calculation leads to the
result (\ref{result}).

\section{Black hole critical exponents from pomeron physics}

In the previous section we have described the kind of strong
gravitational physics that we want to capture using a weakly coupled
gauge theory. Now we look for the appropriate gauge
theory framework to provide a dual description of these phenomena.

The scattering amplitude of two hadrons in the one-pomeron exchange
can be written as \cite{pomeron,books}
\begin{eqnarray}
\mathcal{A}(s,t)=s\int {d^{2}k_{1}d^{2}k_{2}\over (\vec{k}_{1}-\vec{q})^{2}
\vec{k}_{2}^{2}}\Phi_{1}(\vec{k}_{1},\vec{q})
f(\vec{k}_{1},\vec{k}_{2},s,\vec{q})_{\rm BFKL}
\Phi_{2}(\vec{k}_{2},\vec{q})
\label{amplitude}
\end{eqnarray}
where $t\simeq -\vec{q}^{\,\,2}$, $\Phi_{1,2}(\vec{k}_{1,2},\vec{q})$
are the so-called impact factors that encode the information about the
coupling of the pomeron to the colliding hadrons and $\vec{k}_{1}$,
$\vec{k}_{2}$ and $\vec{q}$ are two-dimensional transverse momenta.

In what follows we consider the simplest case of zero momentum
transfer, i.e. very diffractive scattering. This seems to be enough to
capture the dual s-wave gravitational collapse with spherical
symmetry. In the case of zero momentum transfer, $t=0$, and large $s$,
the BFKL kernel in the leading $\log{s}$ approximation and at weak
coupling can be written as
\begin{eqnarray}
f(\vec{k}_{1},\vec{k}_{2},s,\vec{0})_{\rm BFKL}=
{1\over 2\pi^{2}\sqrt{\vec{k}_{1}^{\,2}
\vec{k}_{2}^{\,2}}}\int_{-\infty}^{\infty}d\nu\,e^{\overline{\alpha}_{s}
\chi_{0}(\nu)\log\left({s\over
\vec{k}^{\,2}}\right)+i\nu\log\left({\vec{k}_{1}^{\,2}\over\vec{k}_{2}^{\,2}}\right)}
\end{eqnarray}
where $\overline{\alpha}_{s}=g^{2}N/(4\pi^2)$, $\vec{k}$ is a characteristic
transverse momentum scale and $\chi_{0}(\nu)$ is given in
terms of digamma functions by
\begin{eqnarray}
\chi_{0}(\nu)=2\psi(1)-\psi\left({1\over 2}-i\nu\right)-\psi\left({1\over 2}+i\nu\right).
\end{eqnarray}

In order to establish the holographic map it is convenient to
introduce the quantities
\begin{eqnarray}
y=\overline{\alpha}_{s}\log\left({s\over \vec{k}^{2}}\right), \hspace*{1cm}
\tau_{i}=\log\left({\vec{k}_{i}^{2}\over\vec{k}^{2}}\right) \hspace*{0.5cm} (i=1,2)
\label{y}
\end{eqnarray}
as well as $\eta={1\over 2}-i\nu$. Then we can define 
\begin{eqnarray}
\Psi(y,\tau_{1}-\tau_{2})\equiv 
\pi\vec{k}^{2}_{2}f(\vec{k}_{1},\vec{k}_{2},s,\vec{0})_{\rm BFKL}
=\int_{{1\over 2}-i\infty}^{{1\over 2}+i\infty}{d\eta\over 2\pi
i}e^{\,\,y\chi_{0}(\eta) -\eta(\tau_{1}-\tau_{2})}
\label{Psi}
\end{eqnarray}
where now $\chi_{0}(\eta)=2\psi(1)-\psi(\eta)-\psi(1-\eta)$. The
function $\Psi(y,\tau)$ satisfies the imaginary time Schr\"odinger
equation
\begin{eqnarray}
{\partial\over\partial y}\Psi(y,\tau)=\widehat{H}\Psi(y,\tau),
\label{eom}
\end{eqnarray}
where the Hamiltonian is given by
\begin{eqnarray}
\widehat{H}=\chi_{0}\left(-i{\partial\over\partial\tau}\right).
\label{BFKLham}
\end{eqnarray}
From this expression we see that  
the eigenvalues of this Hamiltonian associated with the plane-wave
eigenfunctions $\psi_{\nu}(\tau)={1\over
\sqrt{2\pi}}e^{i\nu\tau}$ are given by $\chi_{0}(\nu)$. 
Eq. (\ref{eom}) governs the evolution of the scattering amplitude to
higher energies in the BFKL regime. In the holographic map proposed in
Ref. \cite{BPST}, the Hamiltonian (\ref{BFKLham}) is identified with
the Laplacian associated with the $h_{++}$ perturbations of the metric
on the gravity side, with the coordinate $\tau$ identified with the
holographic direction.

The BFKL amplitude (\ref{amplitude}) leads to violations of unitarity
due to its growth at large $s$. In the variables $(y,\tau)$ this
corresponds to an exponential increase of the amplitude with $y$. This
exponential growth is the one {\it we conjecture} to correspond, on
the gravity side, to the growing mode associated to the black hole
formation at threshold. 

In order to extract the critical exponent, we evaluate the integral
in Eq. (\ref{Psi}) using a saddle point approximation. We
calculate the saddle point at a value of $\tau=\tau_{c}(y)$ that
makes the leading exponential equal to one
\cite{mueller,mueller_trian}. Then, the equations determining both the
saturation scale $\tau_{c}(y)$ and the saddle point value $\eta_{c}$
are
\begin{eqnarray}
y\chi_{0}'(\eta_{c})=\tau_{c}(y), \hspace*{1cm} y\chi_{0}(\eta_{c})=
\eta_{c}\tau_{c}(y).
\end{eqnarray}
These equations determine the value of $\eta_{c}$ to be
\begin{eqnarray}
\eta_{c}\chi'_{0}(\eta_{c})-\chi_{0}(\eta_{c})=0, \hspace*{1cm}
\Longrightarrow \hspace*{1cm} \eta_{c}=0.627549,
\end{eqnarray}
while the critical value of $\tau$ as a function of the rapidity $y$
is given by
\begin{eqnarray}
\tau_{c}(y)=\chi_{0}'(\eta_{c})y.
\label{tau-y}
\end{eqnarray}
At this saddle point the function $\Psi(y,\tau)$ becomes
\begin{eqnarray}
\Psi(y,\tau)\approx {e^{-{\tau^2\over 2\chi_{0}''(\eta_{c})y}}\over 
\sqrt{2\pi\chi_{0}''(\eta_{c})y}}e^{\eta_{c}[\tau_{c}(y)-\tau]}.
\end{eqnarray}
A more detailed calculation of the saturation scale $\tau_{c}(y)$ leads 
to logarithmic corrections to the right hand side of Eq. (\ref{tau-y}) but 
does not change the leading linear behavior \cite{mueller_trian}.

Physically speaking the variable $\tau=\tau_{1}-\tau_{2}$ gives the
logarithm of the quotient between the characteristic length scales of
the two colliding hadrons as probed by the pomeron. For a given value
of $\tau$, saturation (and perturbative violations of unitarity)
occurs for a value of the rapidity $y$ such that
$\tau_{c}(y)=\tau$. This value of $y$, where the linear BFKL evolution
breaks down, is given by Eq. (\ref{tau-y}). In terms of the
variables $\vec{k}_{1}$, and $\vec{k}_{2}$ it is
\begin{eqnarray}
e^{y}
=\left({|\vec{k}_{1}|\over |\vec{k}_{2}|}\right)^{2\over 
\chi'_{0}(\eta_{c})}.
\end{eqnarray}
According to the conjecture stated above, the exponent in this
expression should correspond to the Choptuik exponent for critical
black hole formation in five dimensions. A numerical evaluation gives
\begin{eqnarray}
\gamma_{\rm BFKL}={2\over\chi'_{0}(\eta_{c})}=0.409552,
\label{number}
\end{eqnarray}
in remarkable agreement with the gravitational computations as shown in
Eq. (\ref{result}).

Here we have compared the
BFKL value of the critical exponent with the one obtained in numerical 
simulations of gravitational collapse in five-dimensions. As explained in 
\cite{ reviews_choptuik,hara}, the critical exponent can be alternatively computed
as the Liapunov exponent for unstable linear perturbations around the critical solution. 
In the case when this is continuous self-similar, this Liapunov exponent has been computed
in four-dimensions with an accuracy larger than the one provided by numerical simulations. 
Unfortunately, there is to date no analogous computation in five dimensions to compare with
the result of our field theory analysis \cite{fa}. 

\section{Discussion and concluding remarks}

We have presented a prescription to calculate the five-dimensional
Choptuik critical exponent from the QCD pomeron. According to our conjecture,
this critical exponent is determined by the asymptotic value of the 
saturation exponent in QCD in the Regge limit  $\log{(s/\vec{k}^{\,2} )}\rightarrow \infty$ 
at weak coupling, $\overline{\alpha}_s\rightarrow0$, with the variable $y$ 
defined in Eq. (\ref{y}) fixed. Because this double limit is well described by
the leading logarithm approximation, we can ignore the effects of the running of 
the strong coupling constant.

It is important to keep in mind that the value of the critical exponent (\ref{number}) 
provided by the BFKL calculation might be subject to theoretical uncertainties coming 
from saturation and unitarization effects. The situation is
somewhat similar to the one encountered in large-$N$ calculations, where a reliable 
estimation of the size of the corrections to the planar result is difficult to obtain.

The role played by unitarity in the BFKL computation confirms the importance of this
property in black hole dynamics (it has been suggested in a different
context \cite{giddings} that black hole production could lead to a
saturation of the Froissart-Martin bound).

We should stress that the proposed duality is a gravity/CFT
correspondence since, as it is proved in Ref. \cite{lipatovCFT}, there
is a CFT structure encoded in the BFKL Hamiltonian with conformal
weights determining the dependence on the holographic
direction. Moreover, our analysis is robust with respect to
supersymmetry, since supersymmetric effects only show up in the
next-to-leading order
\cite{lipatovSUSY}. This is because fermion loops do not
contribute in the leading logarithm approximation to the diagrams
involved in the BFKL pomeron analysis.  It is also important to keep
in mind that because the BFKL pomeron resums the leading logarithm
contributions to all orders in perturbation theory the result
automatically scales with the 't Hooft coupling
$\overline{\alpha}_{s}\sim g^{2}N$. Therefore it gives the leading
large-$N$ contribution even if we work at finite $N$. Thus, planarity
is generated by the leading $\log{s}$ approximation.

Following the holographic map for pomeron physics suggested in
\cite{BPST} it would be interesting to study the properties of the bulk
critical geometry from the BFKL Hamiltonian at weak coupling. On the
other hand, as a marginal comment, we just mention that one can also try
a naive computation of the critical exponent in the limit 
$\overline{\alpha}_{s}\rightarrow\infty$. 
Using the results of Ref. \cite{stasto} one finds that in this regime 
the saturation condition for the scattering amplitude of two dipoles is given by
\begin{eqnarray}
y-{1\over 2}\log\left({\vec{k}_{1}^{\,2}\over\vec{k}_2^{\,2}}\right)=0
\end{eqnarray}
Repeating our previous analysis, this leads to a value $\gamma_{\rm BFKL}=1$ 
for the saturation exponent at strong coupling. Very likely this 
corresponds to some kind of mean field approximation in critical 
black hole formation. At any rate,
we find this calculation of the Choptuik exponent less compelling than the
one presented for weak coupling.

Apart from the exponent $\gamma$, the solution to the Einstein
equations describing critical gravitational collapse is characterized
by an additional symmetry. It is either discretely or
continuous self-similar, i.e. it has a discrete or continuous
conformal symmetry. We are currently investigating the QCD analog of
this phenomenon, and our findings will be reported elsewhere
\cite{fa}.

\section*{Acknowledgments}

It is a pleasure to thank Rafael Hern\'andez, Elias Kiritsis, Kerstin
Kunze, Marcos Mari\~no and Agust\'{\i}n Sabio Vera for useful
discussions. The work of C.G. has been partially supported by Spanish
DGI contract FPA2003-02877 and CAM grant HEPHACOS
P-ESP-00346. M.A.V.-M. acknowledges partial support from Spanish
Government Grants PA2005-04823 and FIS2006-05319, and thanks the CERN
Theory Group for hospitality and support.

{\sl We would like to dedicate this article to the memory of Pere Pascual. His 
titanic efforts to raise the level of Theoretical Physics in Spain and
his unswerving and uncompromising commitment to the quality of scientific 
research were legendary. His human and scientific presence will be dearly missed. This 
letter is a small token to honor the memory of a great man.}

\end{document}